\documentclass[manuscript]{acmart}

\AtBeginDocument{%
  \providecommand\BibTeX{{%
    \normalfont B\kern-0.5em{\scshape i\kern-0.25em b}\kern-0.8em\TeX}}}

\copyrightyear{Michelle Cohn 2024. This is the author's version of the work. It is posted here for your personal use. Not for redistribution. The definitive Version of Record was published in \textit{Human Factors in Computing Systems, May 11--16 2024}.}
\acmYear{2024}
\acmDOI{XXXXXXX.XXXXXXX}

\acmConference[CHI '24]{Make sure to enter the correct
  conference title from your rights confirmation emai}{May 11--16,
  2024}{Honolulu, HI}
\acmBooktitle{Honolulu '24: Human Factors in Computing Systems,
 May 11--16, 2024, Honolulu, HI} 
\acmISBN{978-1-4503-XXXX-X/18/06}

\usepackage{graphicx}
\usepackage{multirow}
\usepackage{float}

\DeclareUnicodeCharacter{2212}{-}
\begin{document}

\title[Believing Anthropomorphism]{Believing Anthropomorphism: Examining the Role of Anthropomorphic Cues on Trust in Large Language Models}


\author{Michelle Cohn}
\email{mdcohn@ucdavis.edu}
\orcid{0000-0002-4847-1464}
\affiliation{%
  \institution{University of California, Davis}
  \city{Davis}
  \state{California}
  \country{USA}
  \postcode{95616}
}
\affiliation{%
  \institution{Google Research}
  \city{San Francisco}
  \state{California}
  \country{USA}
}
\author{Mahima Pushkarna}
\email{mahimap@google.com}
\orcid{0000-0002-5903-5510}
\affiliation{
  \institution{Google Research}
  \city{Cambridge}
  \state{Massachusetts}
  \country{USA}
}

\author{Gbolahan O. Olanubi}
\email{femio@google.com}
\orcid{0009-0002-8967-1013}
\affiliation{%
  \institution{Google Research}
  \city{New York}
  \state{New York}
  \country{USA}
}

\author{Joseph M. Moran}
\email{joemoran@google.com}
\orcid{0000-0001-9918-592X}
\affiliation{%
  \institution{Google}
  \city{Cambridge}
  \state{Massachusetts}
  \country{USA}
}
\author{Daniel Padgett}
\email{padgettd@google.com}
\orcid{0009-0004-7922-1064}
\affiliation{%
  \institution{Google Research}
  \city{San Francisco}
  \state{California}
  \country{USA}
}

\author{Zion Mengesha}
\orcid{0000-0001-6587-7485}
\email{zmengesh@stanford.edu}
\affiliation{%
  \institution{Stanford University}
  \city{Stanford}
  \state{California}
  \country{USA}
  \postcode{94305}
}
\affiliation{%
  \institution{Google Research}
  \city{San Francisco}
  \state{California}
  \country{USA}
}

\author{Courtney Heldreth}
\email{cheldreth@google.com}
\orcid{0000-0002-9921-7247}
\affiliation{%
  \institution{Google Research}
  \city{Seattle}
  \state{Washington}
  \country{USA}
  \postcode{98103}
}

\renewcommand{\shortauthors}{Cohn et al.}

\begin{abstract}
People now regularly interface with Large Language Models (LLMs) via speech and text (e.g., Bard) interfaces. However, little is known about the relationship between how users anthropomorphize an LLM system (i.e., ascribe human-like characteristics to a system) and how they trust the information the system provides. Participants (n=2,165; ranging in age from 18-90 from the United States) completed an online experiment, where they interacted with a pseudo-LLM that varied in modality (text only, speech + text) and grammatical person (“I” vs. “the system”) in its responses. Results showed that the “speech + text” condition led to higher anthropomorphism of the system overall, as well as higher ratings of accuracy of the information the system provides. Additionally, the first-person pronoun (“I”) led to higher information accuracy and reduced risk ratings, but only in one context. We discuss these findings for their implications for the design of responsible, human–generative AI experiences.
        
\end{abstract}



 

\received{25 January 2024}

\maketitle

\section{Introduction}
Millions of individuals use voice-based interfaces to interact with technology, including voice assistants like Google Assistant, Amazon’s Alexa, and Apple’s Siri, to name a few \cite{ammari2019music}. More recently, people have started engaging with a new conversational agent: large language models (LLMs) such as OpenAI’s ChatGPT\footnote{\url{https://chat.openai.com/}} and Google’s Bard\footnote{\url{https://bard.google.com/}} (for an overview, see \cite{wu2023brief}). LLMs are capable of generating responses that can remain contextually relevant to the user’s input, can be indistinguishable from human-generated responses \cite{ariyaratne2023comparison, huschens2023you}, and can exceed expert-level skills \cite{schwartz2023enhancing, safdari2023personality}(but see also \cite{gao2023comparing}). These advances are likely to increase how people anthropomorphize LLMs, or the tendency to ascribe human-like attributes to a real or imagined system behavior (e.g., progress bars, as ``thinking” behaviors) \cite{fisher1991disambiguating, epley2007seeing}. As prior work has shown a positive relationship between increased anthropomorphism and trust with conversational agents, avatars, and robot systems \cite{hancock2011meta, roesler2021meta, waytz2014mind, kim2012anthropomorphism, seymour2021exploring, rheu2021systematic}, it is possible that a similar positive effect will be at play with LLMs. The aim of the current paper is to test the relationship between anthropomorphism and trust in LLMs by varying the linguistic cues the system produces.

Why does trust in LLMs matter? A growing body of work has shown that LLMs can recreate harmful stereotypes (e.g., by ability in \cite{gadiraju2023wouldn}; race/ethnicity in \cite{dev2020measuring}; gender in \cite{acerbi2023large}), replicating the real-world biases of their training data. Additionally, in cases where there is little training data coverage, LLMs can generate fictitious data (‘hallucinate’) \cite{zellers2019defending}. Indeed, some have raised possible harms of trusting information provided by LLMs, such as misinformation and discrimination \cite{weidinger2021ethical, dubiel2022conversational, dunn2023generative, bender2021dangers}. Therefore, a deeper understanding of the factors that contribute to how users trust an LLM is needed for responsible human-AI experiences.

This paper addresses this gap and is, to our knowledge, the first study to explicitly manipulate two linguistic anthropomorphic cues in an LLM and measure the impact on multiple dimensions of trust. In our study, participants (n=2,165) from the United States engaged with a pseudo-LLM system to elicit information. We manipulated the degree of anthropomorphic cues along two dimensions, Modality (text only or text + speech) and Grammatical Person (first-person, ``I” or third-person ``the system”). Current LLMs, such as Bard and ChatGPT, use ``I” (e.g., ``As a large language model, I…”) and most commonly people engage with them in a text-only interface, but text + speech options are now available as well \cite{Cadenas_2023}.

We examine the impact of these manipulations on overall ratings of anthropomorphism and trust of the system, as well as trial-by-trial ratings of accuracy, risk, and follow-up validation across five areas: health/well-being, career, medications, travel, and cooking. In sum, this work makes the following contributions:

\begin{itemize}
\item We provide experimental evidence that adding a computer generated voice increases both anthropomorphism of LLMs and trust that the information they provide is accurate.
\item We offer evidence that, in some contexts, users rate the information an LLM provides as more accurate and less risky when the system uses the first-person singular pronoun (“I”).
\end{itemize}


\section{Related Work}


\subsection{Anthropomorphism}

People of all ages have shown a propensity to anthropomorphize computers; that is, to ascribe human behaviors to the system \cite{waytz2010sees, epley2007seeing}. Social response theories of technology equivalence ( \cite{Nass_Steuer_Tauber_1994, Nass_Moon_Morkes_Kim_Fogg_1997, Lee_2008} propose that application of social rules from human-human interaction to human-computer interaction are both subconscious and automatic, such as using politeness norms \cite{nass1999people} or applying gender stereotypes \cite{ernst2020gender}. 

\subsubsection{Anthropomorphism and the Voice}
Some work has explicitly manipulated the degree of anthropomorphism by varying the presence or absence of a voice \cite{waytz2014mind, moussawi2021perceptions, qiu2005online}. For example, \cite{waytz2014mind} found that participants who “drove” an autonomous vehicle that had a name (“Iris”) and produced pre-recorded human audio recordings showed stronger anthropomorphism (a composite of 4 questions) and trust than when the system did not have its own voice. \cite{qiu2005online} found that avatars presented with a text-to-speech (TTS) voice increased ratings of trust, relative to conditions with no voice. In a similar vein, \cite{jensen2000effect} found that interacting dyads better cooperated in computer-mediated communication when their speech was generated with TTS, compared to a text-only chat. Furthermore, \cite{moussawi2021perceptions} found that external raters gave higher ratings of anthropomorphism and trust for a conversational interface when it used a TTS voice (relative to no voice).

While often a human voice generates stronger feelings of anthropomorphism than a TTS voice \cite{qiu2009evaluating}, TTS voices still evoke human-like attributes, such as gender \cite{ernst2020gender, Purington_Taft_Sannon_Bazarova_Taylor_2017}, race/ethnicity \cite{holliday2023siri}, and age \cite{Zellou_Cohn_Ferenc_Segedin_2021}.  Additionally, advances in neural TTS mean that computer generated voices ‘sound’ increasingly human-like to listeners \cite{cohn2020perception}. Taken together, these studies suggest that the presence of a voice -- even if computer-generated -- will evoke an anthropomorphism response in the current study.

\subsubsection{Anthropomorphism and Nouns}
While less studied than the voice or visual cues, there is some work showing that anthropomorphism can be reflected in the nouns used to describe a system. In a study of user reviews of Amazon Echo, \cite{Purington_Taft_Sannon_Bazarova_Taylor_2017} found that users who had more social interactions with the Echo (e.g., as a friend or family member) were more likely to also use the third-person singular feminine pronoun, “she”, or the name “Alexa” to describe the system (more human-like), compared to “it” or “Echo” (less human-like).

Fewer studies have explicitly manipulated the nouns the system uses. \cite{nass2000machines} include, ``It doesn’t even refer to itself as ``I,’” as one of the reasons users know a computer is not human-like. While voice assistants have often been developed to avoid the first-person singular (``I”) as a design principle (e.g., Apple’s Siri ``Avoid using personal pronouns”; \cite{siri_guidelines}), current LLMs do use ``I” and ``me” (e.g., ``Tell me what’s on your mind”). While limited, some related work has explored introducing the first-person ‘‘I” pronouns. For example, \cite{qu2022effect} manipulated whether a conversational agent used more first-person singular (``I”) or second-person singular (``you”) pronouns; they found that users preferred statements oriented toward themselves (i.e., that used the second-person). The current study explores the effect of the system using different nouns -- ``I” vs. ``the system” -- in modulating degrees of anthropomorphism and trust, addressing a gap in our scientific knowledge.

\subsection{Trust and Anthropomorphism}
Trust is ``a multifaceted concept that can refer to the belief that another will behave with benevolence, integrity, predictability, or competence” \cite{waytz2014mind}. As expected, the accuracy (either apparent or real) \cite{yin2019understanding} and credibility \cite{metzger2013credibility} of the system improve users’ trust and reliance on it. At the same time, attributes unrelated to performance shape metrics of user trust with avatars, autonomous vehicles, robots, voice assistants, and other conversational agents \cite{hancock2011meta, roesler2021meta, waytz2014mind, kim2012anthropomorphism, seymour2021exploring, rheu2021systematic} (but see \cite{roesler2024dynamics}). For example, \cite{van2019trust} found that a robot that had static eye color (more human-like) was more trustworthy than a robot whose eye color changed according to its status (e.g., green when speaking; blue when it recognized a person). Other work has examined how presence or absence of a human-like face shapes trust \cite{baylor2003effects, gong2008social}. For example, \cite{gong2008social} found users had greater trust for a computer when it was presented with a more human-like face.

As mentioned, the presence of a system’s “voice” has also shown to increase both trust and anthropomorphism in related work, relative to no-voice or text-only conditions \cite{waytz2014mind, moussawi2021perceptions, qiu2005online}. Relatedly, other work has shown that the properties of a system’s voice shape trust:  \cite{torre2018trust} found that participants invested more money in a system that used a Southern Standard British English (SSBE) TTS voice compared to TTS voices that used regionally specific dialects (e.g., Liverpool, Birmingham), but only in the condition where the system gave participants more favorable returns, suggesting that anthropomorphic features interact to shape trust.

\section{Methods}
The driving research question for this experiment was to understand the influence of implicit cues within language and in system feedback design on the extent to which users anthropomorphize the LLM system and trust its outcomes. In large part, our goal was to inform design principles around anthropomorphism for the responsible creation of LLM-powered conversational experiences. 

\subsection{Recruitment and Participants (n=2,170)}
Participants were recruited from a pool of adults from the United States, using a Qualtrics Panel to ensure the participants in each condition were balanced by age, geographic region, and race/ethnicity (see Appendix \ref{tab:demo} for demographic information). To be eligible for the study, all participants had to pass a technical qualifier to ensure that they could hear the sound files\footnote{“Nancy had considered the sleeves.”, generated with a TTS voice (US-English male; “Miles” from TTS Maker; 65 dB); options: "sleeves", "team", "sleds".}, have prior experience with voice technology, have normal hearing (or be wearing a hearing aid), and have normal vision (or be wearing corrective lenses).

 A total of 2,170 US English speaking adults completed the experiment. Three participants were removed as they did not enter a valid age and two participants were removed due to technical issues (n=1 background music, n=1 removed their headphones during the trials and did not hear the audio). Retained participants included data for n=2,165, all of whom indicated they either had normal vision (n=1,300) or that they were currently wearing corrective glasses or contacts (n=865). Participants indicated that they had normal hearing (n=2,100) or were currently wearing hearing aids (n=65). 

Prior to data collection, the survey was reviewed by the Google Ethics Board. The following safety measures were taken to ensure the protection of participants: all participants were given the option to opt out of any sensitive questions (e.g., age, gender, race/ethnicity) and completed informed consent. No identifying information (e.g., name, address, etc.) was collected.

\begin{figure*}[h!]
  \centering
  \includegraphics[width=0.80\textwidth]{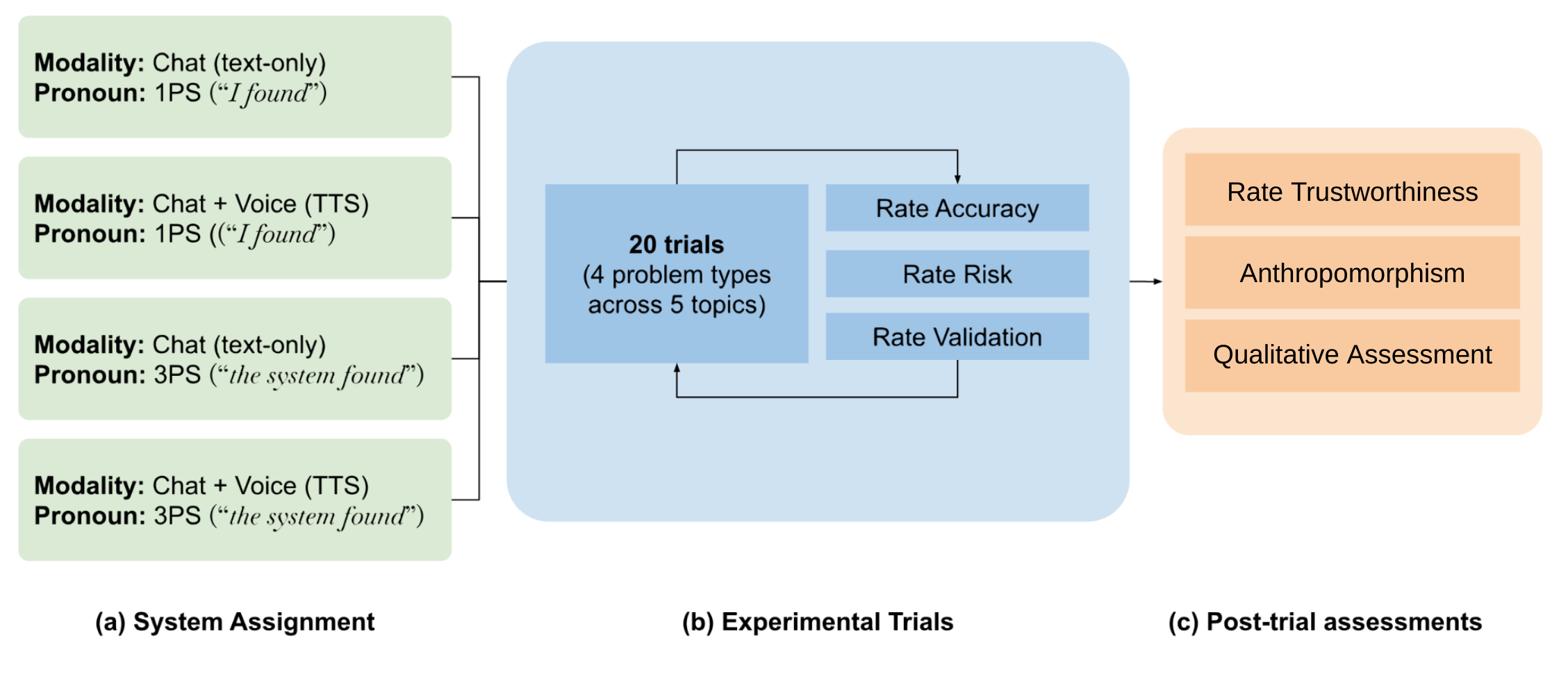}
  \caption{Procedure for the current study. All participants underwent a technical qualifier and comprehension assessment question. \textbf{(a)} System Assignment: First, all qualifying participants were assigned one of four conditional systems in our experiment, controlled for randomization and even distributions. \textbf{(b)} Experimental trials: Participants were asked to complete 20 trials. After each experimental trial, participants rated accuracy, risk, and validation. \textbf{(c)} Post-experimental trial assessments: After completing all experimental trials, participants rated their overall perceptions of the system, including trustworthiness rating, an Anthropomorphism questionnaire, and a qualitative assessment. In this study, we present the results of anthropomorphism and trust assessments.}
  \Description{Experimental procedure}
   \label{fig:procedure}

\end{figure*}

\subsection{Procedure}
\label{section:procedure}

An overview of the experiment is provided in Figure \ref{fig:procedure}. Participants were randomly assigned to one of the four conditions, and were given instructions for the experimental trials (order randomized). On each trial, they asked the system a predefined question and rated the system’s response (see Appendix \ref{appendix:prompts} for full list of stimuli and system responses).

After the system provided a response, participants were given time to review the response in the modality of the system. Participants who experienced the chat (text) modality progressed through the following steps (see Appendix \ref{appendix:trial_schematic} for a schematic): 
    (A) Participants were asked to click an “Ask” button to query the system. 
    (B) A ``submit” sound was used to confirm that the question had been asked after 100ms. 
    (C) Following a 500ms delay, the question was presented in a speech bubble indicating a user-asked question. The system’s response was buffered with a progress cue, in which the user was shown ``...” in the system’s speech bubble for 1500ms. 
    (D) After 2 seconds, the system’s response was displayed in the system’s speech bubble.



Participants who experienced the chat (text) and voice (TTS) modality progressed through steps 1 through 4, with the distinguishing factor being that the system’s response was vocalized using a Google Studio AI voice\footnote{For an audio illustration, see \url{https://cloud.google.com/text-to-speech/docs/voices}.} (US-English Studio-O, female; 65 dB normalized in Praat) at step 4 in addition to the speech bubble. 

In both cases, the system's responses were identical across all participants, while the initial frame varied according to the first-person noun ("I") or third-person noun ("the system") conditions applicable to them ("Here's what [I | the system] found"). After each trial or ``interaction”, participants were asked to rate the system’s response for three dimensions related to trust: 

\begin{enumerate}
    \item \textbf{Perceived Accuracy:} "How accurate is the system's response?" 
    Not at all accurate < Somewhat accurate < Mostly accurate < Completely accurate
    \item \textbf{Perceived Risk}: "If you were relying on this information, how risky would it be if the system got this information wrong?"
    Not at all risky < Neither risky nor unrisky < Somewhat risky < Extremely risky
    \item \textbf{Follow-up Validation:} ``Would you validate the system's answer with other sources?" 
    No < Maybe < Yes
\end{enumerate}



Once participants completed all twenty trials, they were then asked to perform a set of post-trial assessments to rate the system as a whole. 

\begin{itemize}
    \item \textbf{Trustworthiness:} Rate the overall trustworthiness of the system (5 point scale) (Untrustworthy / Trustworthy).
    \item \textbf{Anthropomorphism:} Rate the system on five attributes adapted from the Godspeed Questionnaire \cite{Bartneck_Kulic_Croft_Zoghbi_2009}(all 5 point scales): Authenticity, Humanism, Awareness, Realism, and Competence,
     where each property is out of a possible score of 5 (maximum possible score = 25) (see Appendix \ref{appendix:godspeed}).

\end{itemize}

Finally, participants completed a listening comprehension question\footnote{“They had a problem with the cliff” in a US-English male voice (“Miles” from TTS Maker; 65 dB) and identified the target word ("cliff", "slip", "trip").} to ensure that participants in the chat + voice condition were listening to the audio throughout the experiment. All participants correctly answered the question.

\section{Analysis and Results}

\subsection{Overall Anthropomorphism by Condition}
\begin{figure}[]
  \centering
  \includegraphics[width=0.35\textwidth]{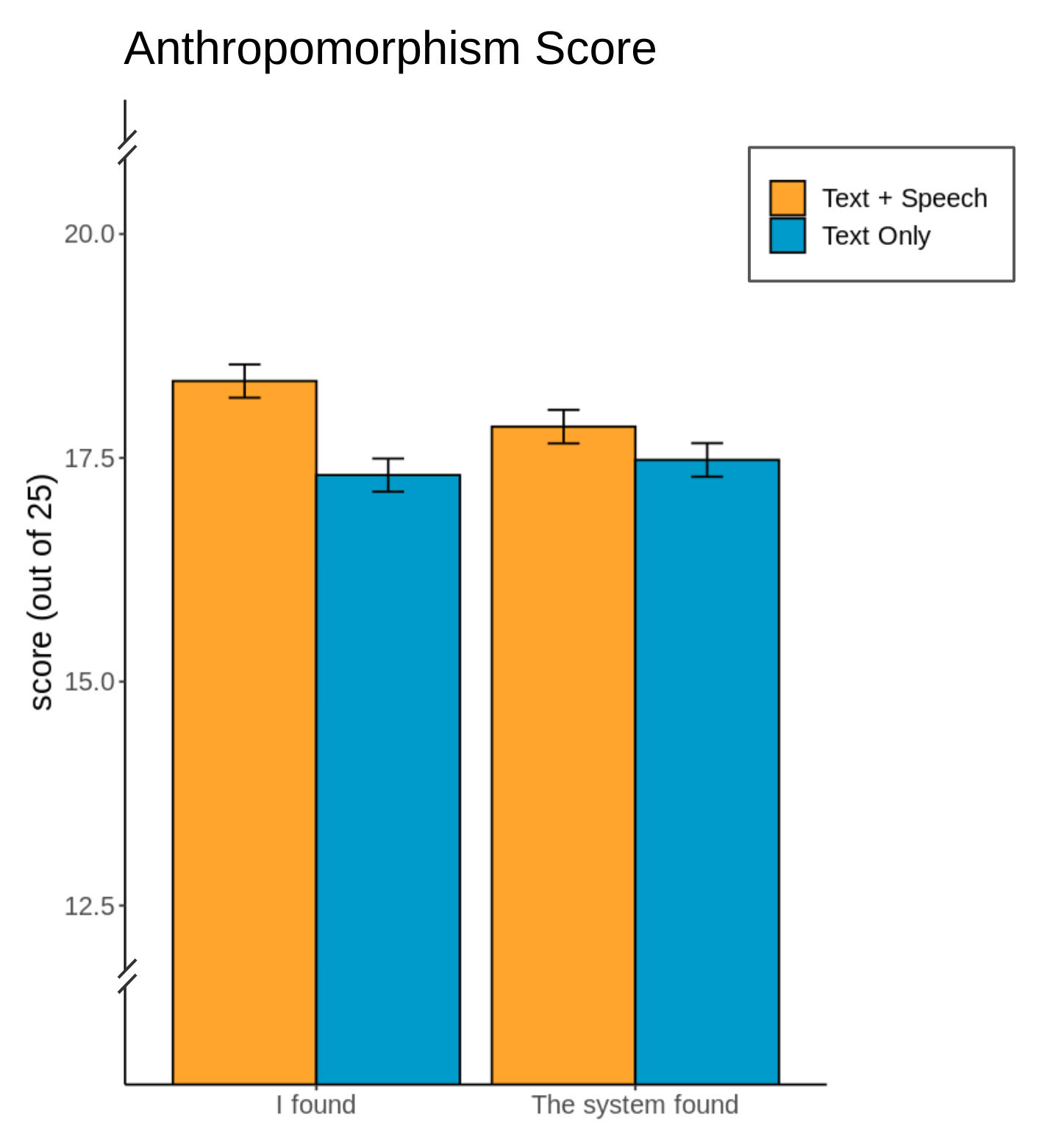}

\caption{Mean anthropomorphism score (out of a total of 25) across Modality (speech + text = orange, text only = dark blue) and Grammatical Person (“I found”, “the system found”). Error bars indicate standard error of the mean.}
  \Description{Shows higher bars for the Speech + Text condition relative to the text only condition, across both the first- and third-person noun conditions for overall anthropomorphism score.}
   \label{fig:anthro}

\end{figure}

We modeled Anthropomorphism Score with a linear regression, with the effects of Modality (speech + text, text only), Grammatical Person (first-person (“I”), third-person (“the system”)), and their interaction. Factors were sum coded.
The model showed no effect of Grammatical Person (first vs. third-person) on anthropomorphism score. As seen in Figure \ref{fig:anthro}, there was an effect of Modality, wherein participants’ anthropomorphism scores were higher when the system responded with both speech and text [Coef = 0.36, t = 3.85, p < 0.001]. There was no interaction between Grammatical Person and Modality.

\subsection{Overall Trustworthiness Rating of System}
\subsubsection{Trustworthiness by Anthropomorphic Manipulations}
We modeled participants’ overall trustworthiness rating of the system with an ordinal regression with the ordinal R package \cite{Christensen_2015}, with the predictors of Modality (speech + text, text only; sum coded), Grammatical Person (first-person, third-person, sum coded), and their interaction. The model showed no effects of Modality or Grammatical Person, or interaction.

\subsubsection{Trustworthiness by Anthropomorphism Score}
We modeled trustworthiness rating with a second ordinal regression to test the overall relationship with Anthropomorphism Score (centered). Results showed individuals with higher anthropomorphism scores rated the system as being more trustworthy [Coef  = 0.26, z  = 23.35, p < 0.001] (see Appendix \ref{appendix:trust_anthro_fig} for a visualization).

\subsection{Experimental Trials: Dimensions of Trust}
We modeled participants’ Likert responses for the three dimensions of trust in the information the system provided (1. Perceived accuracy, 2. Perceived risk, 3. Follow-up validation) with separate ordinal regressions. In each model, fixed effects included Modality (speech + text, text only), Grammatical Person (first-person, third-person), and their interaction. We additionally included predictors of Scenario (health, career, medications, travel, and cooking). Factors were sum coded.

\subsubsection{Information Accuracy}
\begin{figure}[h!]
  \centering
  \includegraphics[width=0.48\textwidth]{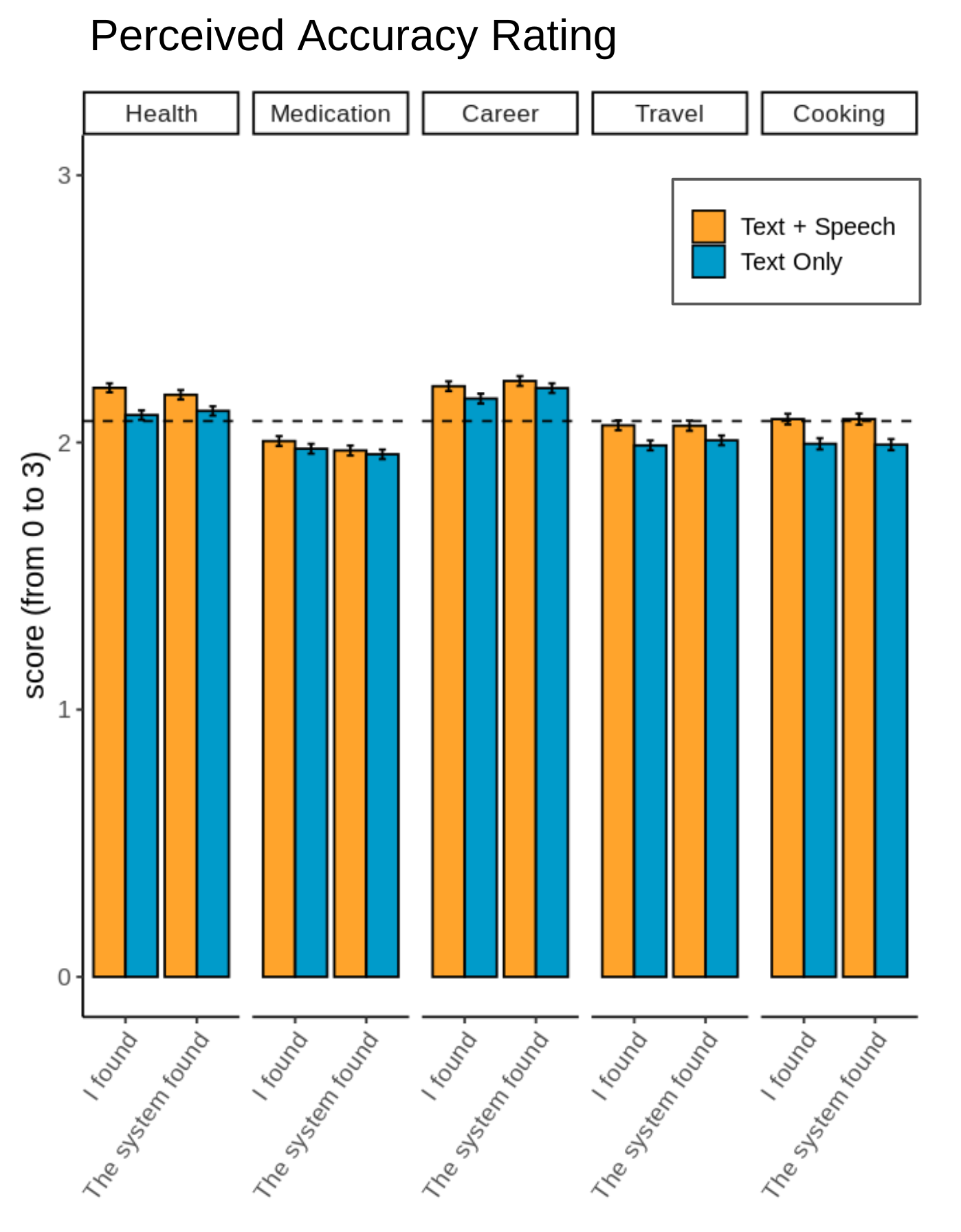}

\caption{Mean ratings for “How accurate is the system’s response?” coded as numeric data (0= “Not at all accurate”, 1 = “Somewhat accurate”, 2 = “Mostly accurate”, 3 = “Completely accurate”) across question contexts. Note that analyses were conducted on the Likert ordinal data and the numeric data is used for visualization purposes only. The experimental conditions included 1) Modality (speech + text = orange, text only = dark blue), and 2) Grammatical Person (“I found”, “the system found”). Error bars indicate standard error of the mean.}
  \Description{Shows higher bars for the speech + text condition for the perceived accuracy of information provided by the system.}
   \label{fig:acc}

\end{figure}

The full model output for accuracy (“How accurate is the system’s response?”) is provided in  Appendix \ref{appendix:accuracy_model}. The model showed an effect of Modality: participants who heard the TTS voice gave higher accuracy ratings (p < 0.001), as seen in Figure \ref{fig:acc}. There was no simple effect of Grammatical Person. There were effects of Scenario: perceived accuracy ratings were higher for the system’s responses for health and career, but lower for medication and travel (all p < 0.001). Interactions between Modality and Scenario revealed reduced accuracy for medication and career responses (both p < 0.001), but higher accuracy for health in the “speech + text” condition (p <0.05). Two interactions between Grammatical Person and Scenario were observed: when using "I", the system’s responses were rated as more accurate for medication but less accurate for career (both p < 0.001). No other interactions were observed.

\subsubsection{Perceived Risk}
The full model output for perceived risk (“If you were relying on this information, how risky would it be if the system got this information wrong?”) is provided in Appendix \ref{appendix:risk_model}. The model showed no simple effects of Modality or Grammatical Person. Scenario was a significant predictor: participants indicated responses about health and medication were riskier to rely on if the system had incorrect information, on average, but scenarios in career and travel were less risky (all p <0.001). Two interactions between Person and Scenario were observed: the systems that used "I" were rated as less risky when giving information about medications, but riskier for careers (both p <0.01). No other effects or interactions were observed.


\subsubsection{Follow-up Validation}
The full model output for validation (“Would you validate the system’s answer with other sources?”) is provided in Appendix \ref{appendix:validate}. The model showed no simple effects of Modality or Grammatical Person. Yet, there were effects for Scenario. Participants indicated they would be more likely to validate the information given for questions about health and medication, but less likely to validate for responses about travel and careers (all p <0.001). One three-way interaction was observed, where participants were more likely to validate responses for travel if the system used “I” and speech + text (p <0.05). No other effects were observed.

\section{Discussion}
This study investigated the role of two linguistic anthropomorphic cues, the presence of a voice and use of the first-person (“I”) by the system, on dimensions of trust in information provided by a pseudo-LLM system. We believe that this is an important step in understanding how people perceive and trust information from LLM models. In the following sections, we describe the five main contributions of this experiment.

\subsection{Linguistic Cues of Anthropomorphism}
First, we found that participants’ overall anthropomorphism scores were higher when the system responded with both text and speech (here a TTS voice), suggesting voice-based effects are not limited to naturally recorded human voices. These findings parallel those in the visual domain, where more human-like qualities tend to increase trust \cite{hancock2011meta, roesler2021meta, waytz2014mind, kim2012anthropomorphism, seymour2021exploring} (but see also \cite{mori2012uncanny}). However, contrary to our prediction, we did not observe effects of the grammatical person (“I” vs. “the system”) on users’ overall anthropomorphism scores. 

\subsection{A Computer Generated Voice Increases Perceived Accuracy of Information}
Second, we found that the presence of a TTS voice shapes perceived accuracy of the system’s response. Relative to a text-only interface, participants believed the information a system gave was more accurate when they also heard the system talk. This effect is robust even in the absence of explicit human-like features, such as an image. This finding is consistent with related work showing that the presence of a pre-recorded human voice (relative to no voice) with an autonomous vehicle system increases both anthropomorphism and trust \cite{waytz2014mind}.

\subsection{First-person (“I”) Increases Dimensions of Trust in Some Contexts}
Third, we also found that, in limited contexts, the first-person "I" shaped dimensions of trust. When the system used "I", its responses were rated as more accurate and less risky for medications; while speculative, these increases could reflect a perception of the system as a domain expert, particularly as the responses contained detailed medication information. 
Conversely, career information was rated as less accurate and riskier to rely on with “I”, raising the possibility that the first-person pronoun could increase perceived subjectivity. Unlike adding a voice, however, introducing “I” had more limited effects in general. One possibility is that participants might have ignored the repeated preamble in each response (“Here’s what [I | the system] found”), whereas the voice was consistent throughout the system’s entire response.

\subsection{Dimensions of Trust are not Equally Affected}
While we observed an impact of the presence of a voice had an across-the-board effect on trial-level ratings of accuracy, we did not observe parallel effects on perceived risk and likelihood to validate. Furthermore, we did not observe that either the modality or person manipulations shaped overall ratings of the system’s trustworthiness, paralleling findings of equal trustworthiness ratings for content generated by humans compared to an LLM \cite{huschens2023you}. 
Additionally, 
we explored several question scenarios that could shape trust. While health and career information were rated as more accurate, medications and travel information were rated as less accurate overall. Participants indicated that systems' responses about health and medications as riskier and ones they were more likely to externally validate. 
Participants’ reticence to rely on medical information given by an LLM reflects the gains needed to improve mutual trust \cite{dunn2023generative}.


\subsection{Implications for UX Responsible Design}
Our findings suggest that people believe information is more accurate and less risky when presented with anthropomorphic cues, which could lead to downstream harms if the system produces non-accurate data \cite{zellers2019defending} or stereotypes \cite{gadiraju2023wouldn, dev2020measuring, acerbi2023large}. We therefore make the following design recommendations: %
\begin{itemize}
    \item Using a voice with an LLM should be limited to cases where the model’s confidence in information accuracy is high. In cases where this cannot be avoided, we suggest introducing cues of speaker uncertainty in the auditory signal \cite{szekely2017synthesising} and source attribution \cite{dunn2023generative}.
    \item Including a generated voice can improve trust and information uptake, which can be used for users’ benefit, such as adherence to a treatment plan in healthcare contexts.
    \item Consider alternatives to first-person pronouns (“I”) in LLMs. While we did not observe a strong overall effect of "I" as for the presence of a voice, it still increased trust in information given about medications.

\end{itemize}

\subsection{Limitations and Future Directions}
While adding the TTS voice had a positive effect on anthropomorphism in the current study, TTS voices can lower ratings of human-likeness, particularly if they are compared to other TTS voices varying in roboticism \cite{Zellou_Cohn_Ferenc_Segedin_2021, Cowan_Branigan_Obregon_Bugis_Beale_2015} or directly to a human voice \cite{Cohn_Ferenc_Segedin_Zellou_2022, gessinger2021phonetic}. Future work directly comparing types of voices can further elucidate these factors for LLMs, as well as reveal any potential uncanniness effects \cite{mori2012uncanny}.
Additionally,
the current study used a female TTS voice, but future research with multiple TTS voices (e.g., varying in gender) can more fully explore the social factors shaping anthropomorphism of LLMs, including probing how their perception might vary across users (e.g., by age, gender, race/ethnicity). Moreover, this study controlled both the question contexts, as well as the exact wording participants “asked”. While this affords experimental control, additional research exploring the effects of a voice and first-person (“I”) will be needed in more naturalistic types of tasks, with real user intents. Finally, our focus was on English in the United States; future work examining other languages and cultural contexts is needed to determine whether effects of anthropomorphic cues and trust for LLMs generalize cross-linguistically and cross-culturally.

\bibliographystyle{ACM-Reference-Format}
\bibliography{sample-base}

\appendix


\clearpage
\onecolumn
\section{Participant Demographics}
\label{tab:demo}
\begin{table}[h]
  \caption{Participant Distribution Across the Four Conditions}
\begin{tabular}{llllll}
\toprule
\multicolumn{2}{l}{}                                                                        & \begin{tabular}[c]{@{}l@{}}First-Person\\ Text only\end{tabular}      & \begin{tabular}[c]{@{}l@{}}First-Person\\ Voice + Text\end{tabular}   & \begin{tabular}[c]{@{}l@{}}Third-Person\\ Text only\end{tabular}      & \begin{tabular}[c]{@{}l@{}}Third-Person\\ Voice + Text\end{tabular}  \\
\midrule
Age            & \begin{tabular}[c]{@{}l@{}}Mean (sd)\\ Range\end{tabular}                  & \begin{tabular}[c]{@{}l@{}}46.9 years (18.3) \\ 18-90\end{tabular} & \begin{tabular}[c]{@{}l@{}}46.7 years (18.4) \\ 18-85\end{tabular} & \begin{tabular}[c]{@{}l@{}}47.1 years (18.0) \\ 18-86\end{tabular} & \begin{tabular}[c]{@{}l@{}}47.1 years (18.5)\\ 18-89\end{tabular} \\
Gender         & Women                                                                      & 280                                                                 & 275                                                                 & 285                                                                 & 281                                                                \\
               & Men                                                                        & 261                                                                 & 261                                                                 & 257                                                                 & 238                                                                \\
               & Another gender                                                             & 3                                                                   & 3                                                                   & 0                                                                   & 1                                                                  \\
Race/ethnicity & white                                                                      & 390                                                                 & 396                                                                 & 396                                                                 & 386                                                                \\
               & \begin{tabular}[c]{@{}l@{}}Hispanic or \\ Latino\end{tabular}              & 96                                                                  & 92                                                                  & 97                                                                  & 99                                                                 \\
               & \begin{tabular}[c]{@{}l@{}}African American \\ or Black\end{tabular}       & 69                                                                  & 69                                                                  & 69                                                                  & 71                                                                 \\
               & \begin{tabular}[c]{@{}l@{}}American Indian\\ or Alaska Native\end{tabular} & 12                                                                  & 10                                                                  & 13                                                                  & 15                                                                 \\
               & Asian American                                                             & 18                                                                  & 24                                                                  & 24                                                                  & 25                                                                 \\
               & \begin{tabular}[c]{@{}l@{}}Hawaiian or \\ Pacific Islander\end{tabular}    & 6                                                                   & 2                                                                   & 2                                                                   & 3                                                                  \\
               & multiracial                                                                & 49                                                                  & 38                                                                  & 38                                                                  & 40                                                                 \\
\multicolumn{2}{l}{\textbf{Total n}}                                                        & \textbf{544}                                                        & \textbf{539}                                                        & \textbf{542}                                                        & \textbf{540}   \\
\bottomrule
\end{tabular}

\end{table}

\clearpage

\section{User Questions and System Responses}
\label{appendix:prompts}

\begin{table}[hbt!]
\caption{User Questions and System Responses across Question Contexts and Types}
\resizebox{\textwidth}{!}{%
\begin{tabular}{lllll}
\toprule
\multicolumn{2}{l}{Question Context}                                                                                                                                       & Question Type & Question                                                                                                               & System Response                                                                                                                                                                                                                                                                                \\
\midrule
\multirow{4}{*}{Medication}  & \multirow{4}{*}{\begin{tabular}[c]{@{}l@{}}You want to learn \\ information about \\ a new medication the \\ doctor mentioned\end{tabular}} & Fact          & \begin{tabular}[c]{@{}l@{}}What’s the dosage for the medication, \\ Levaquin?\end{tabular}                             & \begin{tabular}[c]{@{}l@{}}Here's what [I | the system] found: depending on the type of infection, \\ the usual dose of Levaquin is 250--750 mg given once daily for 5--60 days.\end{tabular}                                                                                             \\
                             &                                                                                                                                             & Risk          & \begin{tabular}[c]{@{}l@{}}Does the medication Corgard have \\ any dangerous side effects?\end{tabular}                & \begin{tabular}[c]{@{}l@{}}Here's what [I | the system] found: side effects for Corgard include: \\ numbness in your hands or feet, dizziness, feeling tired, and upset stomach.\end{tabular}                                                                                           \\
                             &                                                                                                                                             & Advice        & \begin{tabular}[c]{@{}l@{}}Should I take the medication Topamax \\ if I’m already taking Tylenol?\end{tabular}         & \begin{tabular}[c]{@{}l@{}}Here's what [I | the system] found: there are no interactions reported \\ between Topamax and Tylenol. However, this does not necessarily \\ mean no interactions exist. Always consult your healthcare provider.\end{tabular}                               \\
                             &                                                                                                                                             & Estimation    & \begin{tabular}[c]{@{}l@{}}What’s the cost for the medication, \\ Relenza?\end{tabular}                                & \begin{tabular}[c]{@{}l@{}}Here's what [I | the system] found: depending on the pharmacy you \\ visit, the cost for Relenza inhalation powder could be around \$72 for \\ a supply of 20 mg.\end{tabular}                                                                                \\
\multirow{4}{*}{Health}      & \multirow{4}{*}{\begin{tabular}[c]{@{}l@{}}You’re interested in \\ learning more about \\ common health \\ disorders\end{tabular}}          & Fact          & What are symptoms of stroke?                                                                                           & \begin{tabular}[c]{@{}l@{}}Here's what [I | the system] found: common symptoms of stroke include \\ sudden weakness on one side of the body, vision problems, and sudden \\ difficulty speaking.\end{tabular}                                                                           \\
                             &                                                                                                                                             & Risk          & Is cancer hereditary?                                                                                                  & \begin{tabular}[c]{@{}l@{}}Here's what [I | the system] found: up to 10\% of all cancers may be \\ caused by inherited genetic changes. It means that your risk of getting \\ cancer is increased.\end{tabular}                                                                          \\
                             &                                                                                                                                             & Advice        & \begin{tabular}[c]{@{}l@{}}What should I do to reduce my \\ risk of diabetes?\end{tabular}                             & \begin{tabular}[c]{@{}l@{}}Here's what [I | the system] found: to reduce your risk of diabetes, you \\ should eat a balanced diet and exercise regularly.\end{tabular}                                                                                                                  \\
                             &                                                                                                                                             & Estimation    & \begin{tabular}[c]{@{}l@{}}If my dad has heart disease, what are \\ the chances I’ll get it too?\end{tabular}          & \begin{tabular}[c]{@{}l@{}}Here's what [I | the system] found: If one of your immediate family \\ members has had a heart attack, a stroke, or was diagnosed with heart \\ disease before the age of 60, this may indicate a family history of \\ premature heart disease.\end{tabular} \\
\multirow{4}{*}{Career}      & \multirow{4}{*}{\begin{tabular}[c]{@{}l@{}}You’re interested in a \\ career change.\end{tabular}}                                           & Fact          & \begin{tabular}[c]{@{}l@{}}How commonly do people switch jobs \\ into healthcare?\end{tabular}                         & \begin{tabular}[c]{@{}l@{}}Here's what [I | the system] found: it is not as common for people to \\ switch jobs into medicine as it is for other fields, such as computer science.\end{tabular}                                                                                         \\
                             &                                                                                                                                             & Risk          & \begin{tabular}[c]{@{}l@{}}What are the job-related health \\ risks in finance?\end{tabular}                           & \begin{tabular}[c]{@{}l@{}}Here's what [I | the system] found: some risks in finance include chronic \\ stress due to demanding workloads, tight deadlines, and making important \\ financial decisions.\end{tabular}                                                                   \\
                             &                                                                                                                                             & Advice        & \begin{tabular}[c]{@{}l@{}}Would I earn enough as an educational \\ administrator?\end{tabular}                        & \begin{tabular}[c]{@{}l@{}}Here's what [I | the system] found: the average pay range for a Higher \\ Education Administrator varies greatly (as much as \$55,816), which suggests \\ there may be many opportunities for advancement.\end{tabular}                                       \\
                             &                                                                                                                                             & Estimation    & \begin{tabular}[c]{@{}l@{}}How many years of training would I \\ need to do to get into computer science?\end{tabular} & \begin{tabular}[c]{@{}l@{}}Here's what [I | the system] found: a bachelor's program in Computer Science \\ generally takes four years to complete, while a master's degree may take an \\ additional year or two years of coursework.\end{tabular}                                      \\
\multirow{4}{*}{Travel} & \multirow{4}{*}{\begin{tabular}[c]{@{}l@{}}You're interested in \\ learning more about \\ the world.\end{tabular}}                          & Fact          & Where is the world’s tallest tree?                                                                                     & \begin{tabular}[c]{@{}l@{}}Here's what [I | the system] found: the world's tallest tree is Hyperion, which \\ is 344 feet tall and located in Redwood National Park in California in the \\ United States.\end{tabular}                                                                 \\
                             &                                                                                                                                             & Risk          & What are the risks in going in a cave?                                                                                 & \begin{tabular}[c]{@{}l@{}}Here's what [I | the system] found: the most common causes of caving \\ accidents include: falling, being struck by falling objects and hypothermia.\end{tabular}                                                                                            \\
                             &                                                                                                                                             & Advice        & \begin{tabular}[c]{@{}l@{}}Should I visit the world's tallest \\ volcano on a vacation?\end{tabular}                   & \begin{tabular}[c]{@{}l@{}}Here's what [I | the system] found: Mauna Loa is a massive shield volcano. \\ At 13,680 ft above sea level, hiking to its summit requires a good level of \\ physical fitness and endurance.\end{tabular}                                                    \\
                             &                                                                                                                                             & Estimation    & \begin{tabular}[c]{@{}l@{}}How long would it take to reach the \\ top of the world’s tallest mountain?\end{tabular}    & \begin{tabular}[c]{@{}l@{}}Here's what [I | the system] found: from base camp, it takes about 40 days \\ to reach the summit of Mount Everest at 29,029 ft.\end{tabular}                                                                                                                \\
\multirow{4}{*}{Cooking}     & \multirow{4}{*}{\begin{tabular}[c]{@{}l@{}}You're interested in \\ trying a new recipe.\end{tabular}}                                       & Fact          & \begin{tabular}[c]{@{}l@{}}What temperature do I need to cook \\ chicken to?\end{tabular}                              & \begin{tabular}[c]{@{}l@{}}Here's what [I | the system] found: all poultry should reach a safe minimum \\ internal temperature of 165 °F.\end{tabular}                                                                                                                                  \\
                             &                                                                                                                                             & Risk          & \begin{tabular}[c]{@{}l@{}}Can I drink milk that’s been expired \\ for 2 days?\end{tabular}                            & \begin{tabular}[c]{@{}l@{}}Here's what [I | the system] found: as long as it’s been stored properly, \\ unopened milk generally stays good for 5–7 days past its listed date, while\\  opened milk lasts at least 2–3 days past this date.\end{tabular}                                 \\
                             &                                                                                                                                             & Advice        & \begin{tabular}[c]{@{}l@{}}If I have an allergy to peanuts, should \\ I eat peanut oil?\end{tabular}                   & \begin{tabular}[c]{@{}l@{}}Here's what [I | the system] found: a peanut allergy is caused by an allergic\\  reaction to the peanut protein. Peanut oil is typically safe because it's highly \\ refined and has almost no detectable peanut protein.\end{tabular}                       \\
                             &                                                                                                                                             & Estimation    & How much is a dash of salt?                                                                                            & \begin{tabular}[c]{@{}l@{}}Here's what [I | the system] found: a dash is somewhere between 1/16 and \\ 1/8 teaspoon.\end{tabular}     \\                                                        \bottomrule                                                                                         
\end{tabular}%
}

\end{table}

\clearpage
\section{Trial Schematic}
   \label{appendix:trial_schematic}

\begin{figure}[hbt!]
  \includegraphics[width=0.70\textwidth]{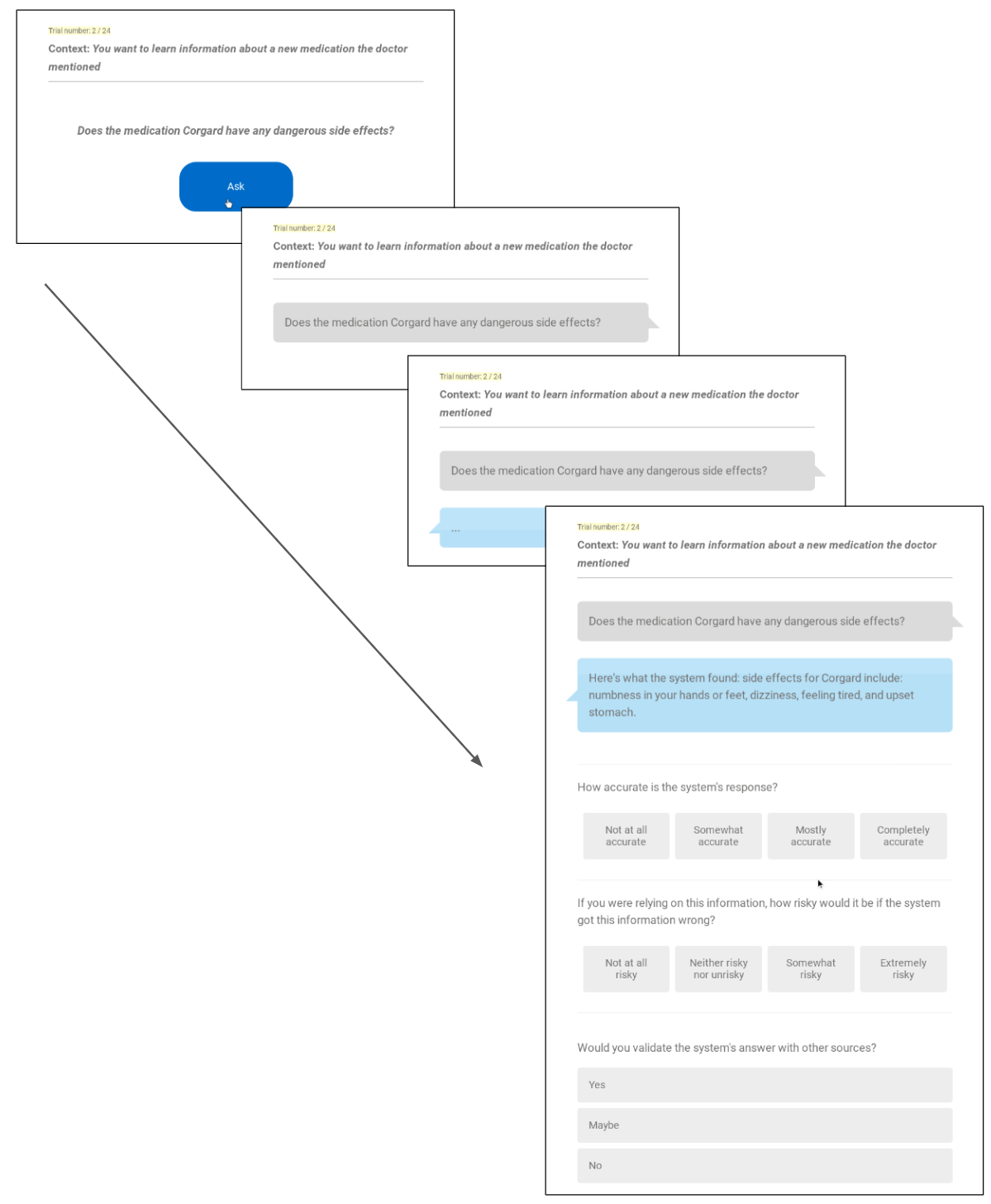}

\caption{First, participants were asked to click an “Ask” button to query the system. A “submit” sound was used to confirm that the question had been asked after 100ms. Following a 500ms delay, the question was presented in a speech bubble indicating a user-asked question. The system’s response was buffered with a progress cue, in which the user was shown “...” in the system’s speech bubble for 1500ms. After 2 seconds, the system’s response was in the system’s speech bubble. In the third-person condition, the response started with “Here’s what the system found”. In the first-person condition, the response started with “Here’s what I found”. In the text-only condition, participants saw the system’s typed response only. In the text + speech condition, they saw the typed response and heard a TTS voice reading the same response aloud. Note that there was a delay after displaying the system’s response before showing the three response options, where participants rated 1) perceived accuracy of the information, 2) perceived risk, and 3) follow-up validation.}
  \Description{This is a schematic of the experimental procedure (see caption)}

\end{figure}
\twocolumn

\clearpage
\section{Anthropomorphism Questionnaire}
\label{appendix:godspeed}
\begin{enumerate}
    \item Please rate your impression of the system: (Fake -- Natural)
    \begin{itemize}
        \item Completely fake
        \item Somewhat fake
        \item Neither fake nor natural
        \item Somewhat natural
        \item Completely natural
    \end{itemize}
    
    \item Please rate your impression of the system: (Machine-like -- Human-like)
    \begin{itemize}
        \item Completely machine-like
        \item Somewhat machine-like
        \item Neither machine-like nor human-like
        \item Somewhat human-like
        \item Completely human-like
    \end{itemize}
    
    \item Please rate your impression of the system: (Unconscious -- Unconscious)
    \begin{itemize}
        \item Completely unconscious
        \item Somewhat unconscious
        \item Neither unconscious nor conscious
        \item Somewhat conscious
        \item Completely conscious
    \end{itemize}
    
    \item Please rate your impression of the system: (Artificial -- Lifelike)
    \begin{itemize}
        \item Completely artificial
        \item Somewhat artificial
        \item Neither artificial nor lifelike
        \item Somewhat lifelike
        \item Completely lifelike
    \end{itemize}    

    \item Please rate your impression of the system: (Incompetent -- Competent)
    \begin{itemize}
        \item Completely incompetent
        \item Somewhat incompetent
        \item Neither incompetent nor competent
        \item Somewhat competent
        \item Completely competent
    \end{itemize}    
    
\end{enumerate}


\clearpage
\onecolumn
\section{Trust and Anthropomorphism Score}
\label{appendix:trust_anthro_fig}
\begin{figure}[h!]
  \centering
  \includegraphics[width=0.45\textwidth]{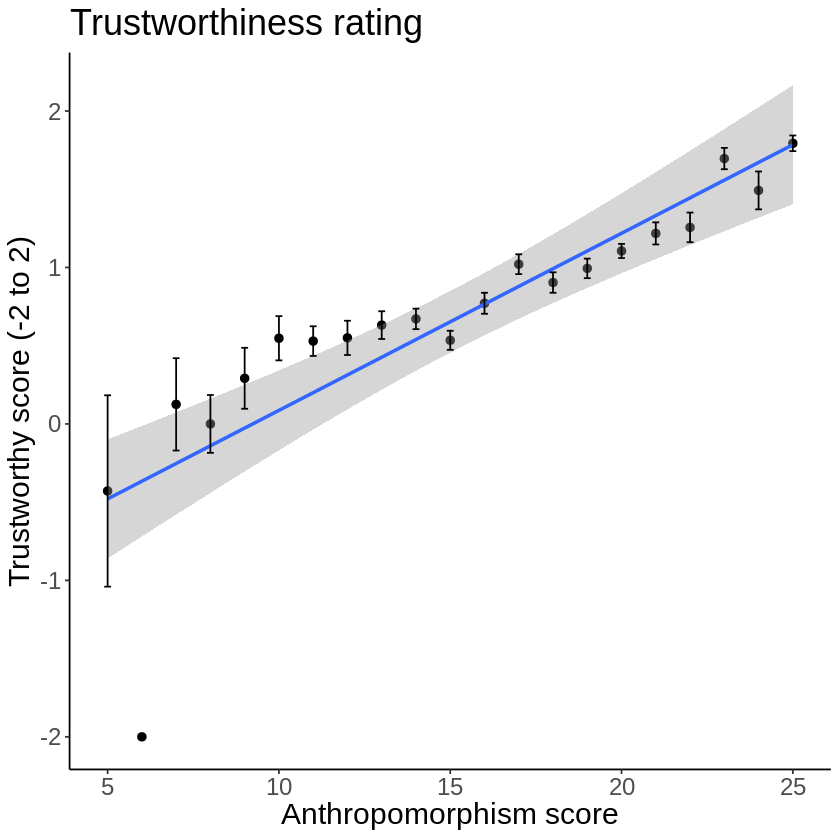}
`  \caption{Relationship between Trustworthiness Rating and Anthropomorphism score. Individual points represent group means at each score. (-2 = Completely untrustworthy, -1 = Somewhat untrustworthy, 0 = Neither untrustworthy or trustworthy, 1 = Somewhat trustworthy, 2 = Completely trustworthy). Error bars indicate standard error of the mean.}
  \Description{This plot shows a positive relationship between a higher anthropomorphism score and trustworthiness.}

\end{figure}

\clearpage
\section{Perceived Information Accuracy}
\label{appendix:accuracy_model}
\begin{table}[hbt!]
\onecolumn
\caption{Ordinal Mixed Effects Model Output for Perceived Information Accuracy}
\begin{tabular}{lrrrr}
\toprule
                                             & \textit{Coef}        & \textit{Std. Error}  & \textit{z value}     & \textit{p value}     \\
\midrule
\textbf{Modality(speech+text)}               & \textbf{0.1}         & \textbf{4.20e-03}    & \textbf{22.95}       & \textbf{<0.001}      \\
Person(“I”)                                  & 5.70e-04             & 4.3e-03              & 0.14                 & 0.89                 \\
\textbf{Scenario(health)}                    & \textbf{0.17}        & \textbf{0.02}        & \textbf{10.47}       & \textbf{<0.001}      \\
\textbf{Scenario(medication)}                & \textbf{-0.32}       & \textbf{4.5e-03}     & \textbf{-71.46}      & \textbf{<0.001}      \\
\textbf{Scenario(career)}                    & \textbf{0.36}        & \textbf{4.5e-03}     & \textbf{79.73}       & \textbf{<0.001}      \\
\textbf{Scenario(travel)}                    & \textbf{-0.16}       & \textbf{4.5e-03}     & \textbf{-35.71}      & \textbf{<0.001}      \\
Modality(speech):Person(“I”)                 & 0.01                 & 4.3e-03              & 1.2                  & 0.23                 \\
\textbf{Modality:Scenario(health)}           & \textbf{0.04}        & \textbf{0.02}        & \textbf{2.25}        & \textbf{0.02}        \\
\textbf{Modality:Scenario(medication)}       & \textbf{-0.05}       & \textbf{4.5e-03}     & \textbf{-12.1}       & \textbf{<0.001}      \\
\textbf{Modality:Scenario(career)}           & \textbf{-0.03}       & \textbf{4.5e-03}     & \textbf{-6.14}       & \textbf{<0.001}      \\
Modality:Scenario(travel)                    & 4.20e-03             & 4.5e-03              & 0.95                 & 0.34                 \\
Person:Scenario(health)                      & 0.01                 & 0.02                 & 0.42                 & 0.68                 \\
\textbf{Person:Scenario(medication)}         & \textbf{0.05}        & \textbf{4.5e-03}     & \textbf{10.68}       & \textbf{<0.001}      \\
\textbf{Person:Scenario(career)}             & \textbf{-0.04}       & \textbf{4.5e-03}     & \textbf{-8.14}       & \textbf{<0.001}      \\
Person:Scenario(travel)                      & -0.01                & 4.5e-03              & -1.79                & 0.07                 \\
Modality:Person:Scenario(health)             & 0.01                 & 0.02                 & 0.85                 & 0.4                  \\
Modality:Person:Scenario(medication)         & 1.90e-03             & 4.5e-03              & 0.43                 & 0.67                 \\
Modality:Person:Scenario(career)             & -1.40e-03            & 4.5e-03              & -0.32                & 0.75                 \\
Modality:Person:Scenario(travel)             & 0.01                 & 4.5e-03              & 1.29                 & 0.2                  \\
\midrule
\textit{Threshold coefficients}              & \multicolumn{1}{l}{} & \multicolumn{1}{l}{} & \multicolumn{1}{l}{} & \multicolumn{1}{l}{} \\
Not at all accurate|Somewhat accurate        & -3.99                & 0.03                 & -156.26              & \multicolumn{1}{l}{} \\
Somewhat accurate|Mostly accurate            & -1.56                & 2.50e-02             & -365.98              & \multicolumn{1}{l}{} \\
Mostly accurate|Completely accurate & 0.6                  & 0.02                 & 38.33                & \multicolumn{1}{l}{} \\
\midrule
\textit{Random effects}                      & Variance             & \multicolumn{1}{l}{} & \multicolumn{1}{l}{} & \multicolumn{1}{l}{} \\
Participant                                  & 2.44                 & \multicolumn{1}{l}{} & \multicolumn{1}{l}{} & \multicolumn{1}{l}{} \\
\bottomrule  
\end{tabular}
\end{table}


\clearpage
\section{Perceived Risk Model}
\label{appendix:risk_model}
\begin{table}[hbt!]
\onecolumn
\caption{Ordinal Mixed Effects Model Output for Perceived Risk}
\begin{tabular}{lrrrr}\\
\toprule
                                             & \textit{Coef}     & \textit{Std. Error} & \textit{z value} & \textit{p value} \\
\midrule
Modality(speech+text)               & -0.03    & 0.03       & -1.12   & 0.26    \\
Person(“I”)                                  & -3.8e-03          & 0.03                & -0.12            & 0.90             \\
\textbf{Scenario(health)}                    & \textbf{0.44}     & \textbf{0.02}       & \textbf{24.35}   & \textbf{<0.001}  \\
\textbf{Scenario(medication)}                & \textbf{0.53}     & \textbf{0.02}       & \textbf{29.31}   & \textbf{<0.001}  \\
\textbf{Scenario(career)}                    & \textbf{-0.20}    & \textbf{0.02}       & \textbf{-10.46}  & \textbf{<0.001}  \\
\textbf{Scenario(travel)}                    & \textbf{-0.99}    & \textbf{0.02}       & \textbf{-52.95}  & \textbf{<0.001}  \\
Modality(speech):Person(“I”)                 & -0.02             & 0.03                & -0.60            & 0.55             \\
Modality:Scenario(health)          & -2.6e-03 & 0.02       & -0.14   & 0.89    \\
Modality:Scenario(medication)       & 1.1e-03  & 0.02       & 0.06    & 0.95    \\
Modality:Scenario(career)           & -1.0e-02 & 0.02       & -0.53   & 0.59   \\
Modality:Scenario(travel)                    & -0.01             & 0.02                & -0.63            & 0.53             \\
Person:Scenario(health)                      & 0.02              & 0.02                & 1.07             & 0.29             \\
\textbf{Person:Scenario(medication)}         & \textbf{-0.05}    & \textbf{0.02}       & \textbf{-2.84}   & \textbf{<0.01}   \\
\textbf{Person:Scenario(career)}             & \textbf{0.05}     & \textbf{0.02}       & \textbf{2.92}    & \textbf{<0.01}   \\
Person:Scenario(travel)                      & -0.02             & 0.02                & -1.32            & 0.19             \\
Modality:Person:Scenario(health)             & 0.01              & 0.02                & 0.46             & 0.64             \\
Modality:Person:Scenario(medication)         & 0.01              & 0.02                & 0.58             & 0.56             \\
Modality:Person:Scenario(career)             & -3.6e-03          & 0.02                & -0.19            & 0.85             \\
Modality:Person:Scenario(travel)             & -0.01             & 0.02                & -0.53            & 0.60             \\
\midrule
\textit{Threshold coefficients}              &                   &                     &                  &                  \\
Not at all risky | Neither risky nor unrisky & -1.52             & 0.03                & -46.82           &                  \\
Neither risky nor unrisky | Somewhat risky   & -4.2e-03          & 0.03                & -0.13            &                  \\
Somewhat risky | Extremely risky    & 1.70              & 0.03                &                  &                  \\
\midrule
\textit{Random effects}                      & Variance          &                     &                  &                  \\
Participant                                  & 1.91              &                     &                  &                 
\\  \bottomrule        
\end{tabular}
\end{table}


\clearpage
\section{Follow-up Validation Model}
\label{appendix:validate}

\begin{table}[hbt!]
\onecolumn
\caption{Ordinal Mixed Effects Model Output for Follow-up Validation}
\begin{tabular}{lllll}\\
\toprule
                                          & \textit{Coef}     & \textit{Std. Error} & \textit{z value} & \textit{p value} \\
\midrule
Modality(speech+text)            & -0.03    & 0.04       & -0.74   & 0.46    \\
Person(“I”)                               & -0.03             & 0.04                & -0.67            & 0.49             \\
\textbf{Scenario(health)}                 & \textbf{0.47}     & \textbf{0.22}       & \textbf{21.68}   & \textbf{<0.001}  \\
\textbf{Scenario(medication)}             & \textbf{0.70}     & \textbf{0.02}       & \textbf{31.04}   & \textbf{<0.001}  \\
\textbf{Scenario(career)}                 & \textbf{-0.34}    & \textbf{0.02}       & \textbf{-16.15}  & \textbf{<0.001}  \\
\textbf{Scenario(travel)}                 & \textbf{-0.48}    & \textbf{0.02}       & \textbf{-23.33}  & \textbf{<0.001}  \\
Modality(speech):Person(“I”)              & 6.7e-03           & 0.04                & 0.17             & 0.87             \\
Modality:Scenario(health)        & -9.3e-03 & 0.02       & -0.43   & 0.67    \\
Modality:Scenario(medication)    & 0.01     & 0.02       & 0.49    & 0.62    \\
Modality:Scenario(career)        & 0.02     & 0.02       & 0.65    & 0.52    \\
Modality:Scenario(travel)                 & -0.01             & 0.02                & -0.56            & 0.57             \\
Person:Scenario(health)                   & 0.04              & 0.02                & 1.89             & 0.06             \\
Person:Scenario(medication)      & 9.1e-03  & 0.02       & 0.41    & 0.68    \\
Person:Scenario(career)          & -0.03    & 0.02       & -1.66   & 0.10    \\
Person:Scenario(travel)                   & -0.02             & 0.02                & -1.15            & 0.25             \\
Modality:Person:Scenario(health)          & 2.03-03           & 0.02                & 0.09             & 0.93             \\
Modality:Person:Scenario(medication)      & -0.01             & 0.02                & -0.48            & 0.63             \\
Modality:Person:Scenario(career)          & -0.02             & 0.02                & -1.03            & 0.30             \\
\textbf{Modality:Person:Scenario(travel)} & \textbf{0.04}     & \textbf{0.02}       & \textbf{2.01}    & \textbf{0.04}    \\
\midrule
\textit{Threshold coefficients}           &                   &                     &                  &                  \\
No | Maybe                                & -2.12             & 0.04                & -49.88           &                  \\
Maybe | Yes                               & -0.33             & 0.04                & -7.95            &                  \\
\midrule
\textit{Random effects}                   & Variance          &                     &                  &                  \\
Participant                      & 3.36              &                     &                  &                  \\

\bottomrule
\end{tabular}
\end{table}

\end{document}